\PassOptionsToPackage{table}{xcolor}
\documentclass[format=acmsmall, review=false, screen=true, anonymous=false]{acmart}

\usepackage{booktabs} 
\usepackage[ruled]{algorithm2e} 
\usepackage{natbib}
\usepackage{enumitem}
\usepackage{tabularx}
\usepackage{siunitx}
\usepackage{multirow}
\usepackage[justification=centering]{caption}
\usepackage[all]{nowidow}
\newcolumntype{Y}{>{\centering\arraybackslash}X}

\SetAlFnt{\small}
\SetAlCapFnt{\small}
\SetAlCapNameFnt{\small}
\SetAlCapHSkip{0pt}
\IncMargin{-\parindent}

\acmJournal{PACMHCI}
\acmVolume{2}
\acmNumber{CSCW}
\acmArticle{81}
\acmYear{2018}
\acmMonth{11}
\acmPrice{15.00}
\acmDOI{10.1145/3274350}

\setcopyright{acmlicensed}

\received{April 2018}
\received[revised]{July 2018}
\received[accepted]{September 2018}

\begin{document}
\title[Tending Unmarked Graves]{Tending Unmarked Graves: Classification of Post-mortem Content on Social Media}

\author{Jialun "Aaron" Jiang}
\affiliation{
  \institution{University of Colorado Boulder}
  \department{Department of Information Science}
  \streetaddress{ENVD 201, 1060 18th St.}
  \city{Boulder}
  \state{CO}
  \postcode{80309}
  \country{USA}}
\email{aaron.jiang@colorado.edu}

\author{Jed R. Brubaker}
\affiliation{
  \institution{University of Colorado Boulder}
  \department{Department of Information Science}
  \streetaddress{ENVD 201, 1060 18th St.}
  \city{Boulder}
  \state{CO}
  \postcode{80309}
  \country{USA}}
\email{jed.brubaker@colorado.edu}

\renewcommand{\shortauthors}{J. A. Jiang \& J. R. Brubaker}

\begin{abstract}
User-generated content is central to social computing scholarship. However, researchers and practitioners often presume that these users are alive. Failing to account for mortality is problematic in social media where an increasing number of profiles represent those who have died. Identifying mortality can empower designers to better manage content and support the bereaved, as well as promote high-quality data science. Based on a computational linguistic analysis of post-mortem social media profiles and content, we report on classifiers developed to detect mortality and show that mortality can be determined after the first few occurrences of post-mortem content. Applying our classifiers to content from two other platforms also provided good results. Finally, we discuss trade-offs between models that emphasize pre- vs. post-mortem precision in this sensitive context. These results mark a first step toward identifying mortality at scale, and show how designers and scientists can attend to mortality in their work.
%
%
%
%
%
\end{abstract}

%
%


%
%


\maketitle


\section{Introduction}

Death is becoming a common part of our online social lives \cite{walter_does_2012}. 
It was recently estimated that over 1.7 million Facebook users will die in 2018 \cite{carroll_1.7_2018}, adding to the growing number of ``post-mortem profiles'' \cite{brubaker_we_2011} that inhabit our social networks and online communities. 
However, encounters with post-mortem profiles and memorial content at unexpected moments can be unsettling. Previous work has shown that while post-mortem profiles and online memorials are important to people,  it can be confusing when death and grief appear in the context of people's everyday social media use \cite{brubaker_beyond_2013}. This type of uncomfortable experience is exemplified by Eric Meyer's widely circulated experience with Facebook's Year in Review---the feature that generates a video with highlights from the previous year. While other people shared video compilations of vacations, parties, and birthdays, all accompanied by joyful and upbeat music, the celebratory video waiting for Mr. Meyer prominently featured the death of his six-year old daughter. For Meyer, the encounter was nothing short of ``cruel'' \cite{meyer_inadvertent_2014}. These upsetting experiences call for ways to detect the mortality of online content so that platform designers can design sensitive interactions with death and bereavement in mind.


However, determining if a user has died, let alone identifying memorial content that exists outside of the context of a user profile, is more difficult than it might seem. Identifying post-mortem accounts remains a manual process \cite{brubaker_legacy_2016}, and most platforms rely on survivors to report the deaths of users. Not only is manual reporting labor intensive and emotionally taxing, depending on survivors to report deaths results in delayed, inconsistent, and ultimately unreliable data.
%
%

As post-mortem profiles and expressions of grief become increasingly common on social media, reliable identification of both also becomes increasingly important. Many platforms may wish to sensitively design for end-of-life and post-mortem scenarios. Facebook, for example, supports the memorialization of profiles as well as post-mortem management options \cite{brubaker_legacy_2016}. However, in the absence of a clear and reliable indicator of mortality, platform designers are constrained. Automatic identification can empower designers in how they handle post-mortem profiles and content. Moreover, timely identification is crucial to supportive and compassionate platform design as it allows designers to address issues such as the dissemination of the news of death, which can potentially soften the impact of such news in public social media spaces, as encounters with such sensitive disclosures can be emotionally draining \cite{andalibi_responding_2018}.
%
%

In this work, we provide a linguistic description of post-mortem content on social media and introduce machine learning-based identification methods for both post-mortem profiles and memorial comments. Using 870,326 comments from 2,688 public profiles on MySpace as training data, we compare the performance of a series of supervised machine learning classifiers on the level of user profiles and individual comments on those profiles with three different feature sets, and show we are able to accurately classify both post-mortem user profiles and single post-mortem comments. Based on an accurate profile classification model, we then show that we are able to quickly identify post-mortem profiles after the users' death, and that our classifiers can also accurate classify death on two other datasets recently collected from Facebook Memorial Groups and newspaper obituaries, which represent significantly different platforms and contexts. We finally discuss the generalizability and applicability of our classifiers to other areas in CSCW, and caution platform designers to carefully consider the trade-offs and consequences when deploying such classifiers.

%
%
%
%
Our results mark a first step toward automatic identification of post-mortem profiles and memorial comments on social media. Automatic identification support new ways to identify and verify post-mortem profiles, enabling data scientists to account for mortality in their datasets, and enabling designers means to provide for granular survivor-support and compassionate platform design.
%
%

\section{Related Work}
CSCW and related scholars have begun to use large scale data science and machine learning approaches to understand and detect behavior online. Notable among this work are studies related to health and wellness, from depression \cite{choudhury_predicting_2013} to postpartum emotion \cite{de_choudhury_predicting_2013} to influenza \cite{huang_examining_2017}. Research successful in using machine learning to predict behaviors often suggests key design recommendations. De Choudhury et al.'s work, for example, focused on efforts to preemptively detect signals of depression based on social media data \cite{choudhury_predicting_2013}; Chancellor et al.'s work on predicting mental illness severity in online pro-eating disorder communities suggested that platform designers could build early-warning systems to provide appropriate and timely support \cite{chancellor_quantifying_2016}. This work points to ways that classification could sensitize the design of platforms and interactions. In this paper we follow in their footsteps by taking machine learning-based approach to detect death and bereavement.

\subsection{Challenges of Post-mortem Content}
The presence of post-mortem accounts in otherwise living networks can be challenging for both social network site users and designers. Designing for post-mortem accounts requires both technical and social considerations. As interaction design on social media platforms increasingly relies on algorithmic content curation, the inability to accurately and efficiently determine the mortality of users becomes a challenge. In the past, highly visible mistakes have been made, for example, when Facebook encouraged users to interact with their inactive Facebook friends, unintentionally including the dead \cite{brubaker_beyond_2013}. More recently, algorithmic curation was at fault in an experience shared by Eric Meyer \cite{meyer_inadvertent_2014}, whose recently deceased daughter was featured prominently in his Facebook Year in Review.

Currently, most social media platforms rely on people to report the death of a user. On most sites, reports result in the deactivation and removal of the account. However, some platforms allow profiles to remain as part of the network. Facebook, for example, memorializes profiles after they learn of person's death. Moreover, Facebook allows people to designate a legacy contact who can make some changes to the deceased's profile in order to be supportive of the bereft community \cite{brubaker_legacy_2016}.

In contrast, Google implemented an inactivity-based approach with Inactive Account Manager \cite{google_about_2015}. Google allows users to specify trusted contacts and will notify them when the account is deemed inactive for longer than a duration chosen by the account owner (the default is three months). Account owners can also elect to allow access to various parts of their account data upon inactivity. While death is a clear use case for Inactive Account Manager, accounts become inactive for a variety of different reasons. As a result, even though inactivity provides a conservative signal by which to identify those who might be dead, it cannot provide a means for positively identifying people who have died.

These common approaches to identifying post-mortem accounts present three problems for designers looking to sensitize their platforms. First, relying on people to report deaths results in inconsistent and unreliable data. Second, while inactivity-based approaches, such as those used by Google, produce more consistent data, they cannot reliably be used as an indicator of death, as people may be inactive for any number of reasons. Finally, any signal from an inactivity-based approach is available only after some period of time (typically months), limiting the ability to design for death during the critical time frame immediately following a death.

The need for sensitivity around post-mortem content, however, is not limited to post-mortem profiles. There are many scenarios in which memorial content is not associated with a profile, but for which sensitivity should still be designed. For example, the Year In Review photo that Eric Meyer saw was not linked to his daughter's profile,  but was nevertheless extremely distressing. Gach et al. also found that disagreement and argument about how one should grieve, namely ``grief policing,'' could result from the different ways people mourn deceased celebrities on Facebook, who did not have Facebook profiles. These problems existing outside the context of profiles point to the need for not only identifying death of a profile, but also of standalone content such as individual comments that may appear independently from the profiles where they belong. 
\subsection{Classifying Mortality}
To provide a solution to these problems, in this paper we make use of machine learning approaches based on computational linguistics to identify post-mortem profiles and content. To our knowledge, no existing work has attempted to classify mortality within user generated online data, with one exception: Ma et al. \cite{ma_write_2017} classified deceased member sites on CaringBridge, an online health community. However, rather than seeking to support platform designers, Ma and his colleagues found that classifying death was a necessary step in their larger data science project---a situation we argue deserves broader attention,  which we discuss in this work.

While Ma et al. were successful at classifying blogs that were relinquished and blogs that represent deceased patients in an online health community to show how separating the deceased can impact data analysis results, here we took a step further to classify the mortality of profiles \emph{and} individual comments on a general social media, where the style of communication is more diverse than that in health communities. This difference in contexts is important also because of the difference in user expectations: while death can be an inevitable topic on a site that logs patients' medical journeys, it comes up much less often on a general social media that tends to be more casual. Therefore, it is reasonable to assume that coming across post-mortem content on a general social media would be considerably more unexpected and distressing, and our work on classifying death in this context has important implications for designing general social media platforms.
%
%

\subsection{Death on the Internet}
%
%
%
%
%
%
To situate our analysis, it is important to consider how death is memorialized online and the impacts that the Internet is having on the ways we experience grief. The Internet, and social media specifically, has drawn both private and public aspects of grief into everyday social media life \cite{walter_does_2012}. Social media easily extends who can participate or witness in these practices beyond the family, while also providing a public audience for what has historically been private communication with the dead \cite{walter_does_2012}. 

A growing body of research in social computing and related fields has documented how people connected to the deceased continue to interact with post-mortem profiles 
\cite{brubaker_we_2011,brubaker_beyond_2013,getty_i_2011,lingel_digital_2013,willis_mourning_2017}, 
reappropriating them into digital memorials 
\cite{mori_design_2012,pitsillides_digital_2009},
personal archives \cite{acker_death_2014,lingel_digital_2013,brubaker_stewarding_2014}, 
and gathering places for online communities \cite{pitsillides_digital_2009,getty_i_2011,brubaker_beyond_2013,willis_mourning_2017}. 
In terms of content, comments posted soon after death express shock and disbelief, as well as acknowledging the death, often with the phrase ``RIP.'' Reading across the literature, engagement with the deceased falls in four major categories: sharing memories, expressing condolences, posting updates, and maintaining connections with both the deceased and their network \cite{brubaker_select_2011,willis_mourning_2017}. Collectively, this scholarship indicates that post-mortem content is marked by distinctive linguistic characteristics, showing promise in using language to identify post-mortem content.
%
%
%
%
%

\subsection{The Language of Death and Bereavement}
Previous research has studied language use around death using computational methods in a variety of contexts, from classifying reports of death \cite{imane_multi-label_2017} to identifying suicidal signals  \cite{mulholland_suicidal_2013,desmet_emotion_2013,desmet_recognising_2014,pestian_suicide_2010,huang_exploring_2017}, and has achieved promising results. Research of bereaved language has also revealed important patterns of sentiment and linguistic style. 

Researchers have often found success using computational linguistics tools like LIWC \cite{tausczik_psychological_2010}. In comments made to post-mortem social media profiles, words from LIWC's first person singular pronoun, past tense verb, adverb, preposition, conjunction, and negation categories appear more frequently in emotion distressed (ED) comments; in terms of sentiment, ED comments also showed higher use of anger words \cite{brubaker_grief-stricken_2012} . These results suggest highly emotional bereaved language are accompanied with distinctive linguistic signals. Likewise, in Ma et al.'s \cite{ma_write_2017} study of writing on an online health community, the they were able to use journals (blog-like entries that included obituaries and funeral announcements) alongside bereaved content to successfully classify mortality using words from LIWC's death dictionary (e.g., died, funeral, grave). Together, these studies show great potential in using computational linguistics to analyze distinctive changes in language after death, and served as inspiration for our work classifying the mortality of social media content.

\section{Data \& Prior Analysis}
The data used for the work we present in this paper are a subset of comments collected from public social media profiles. Complete details on how this dataset was constructed, as well as details about the corpora can be found in our previous work \cite{ICWSM1817819}. We provide a description of this dataset here to situate the analysis and results presented in this paper.

Our dataset consists of the content posted to publicly visible MySpace profiles of people who lived in the United States during the three years following their death.  Deceased profiles were identified using MyDeathSpace (MDS), a website dedicated to connecting obituaries and/or news of deaths to existing MySpace profiles. The dataset was collected in April 2010, and to verify mortality, each profile was hand-checked by one of the researchers using comments posted to the profile or changes made to the profile itself (e.g., adding ``RIP'' or ``In memory of'' to the profile name). 
The dataset consists of 870,326 comments from 2,688 unique profiles. Descriptive statistics of the dataset are shown in Table \ref{tab:1}. 

\begin{table}%
\caption{Descriptive statistics of the final training dataset from MySpace.}
\label{tab:1}
\begin{minipage}{\columnwidth}
\begin{center}
\begin{tabular}{ll}
  \toprule
  Total comments & 870,326 \\
  Total profiles    & 2,688\\
  Post-mortem comments  & 324,089 (37.24\%)\\
  Pre-mortem comments     & 546,327 (62.76\%)\\
  Average comments per profile    & 323.78\\
  Median comments per profile   & 169.50\\
  Average words per comment       & 32.74\\
  Average post-mortem comments per profile     & 120.57\\
  Average per-mortem comments per profile & 203.21\\
  \bottomrule
\end{tabular}
\end{center}
\end{minipage}
\end{table}


\begin{table}%
\caption{Top 5 frequently occurring unigrams and bigrams in pre- and post-mortem comments.}
\label{tab:2}
\begin{center}
\begin{tabular}{cccccccc}
	\multicolumn{4}{c}{Unigram} & \multicolumn{4}{c}{Bigram} \\
    \cmidrule(lr{.75em}){1-4} \cmidrule(lr{.75em}){5-8}
	\textbf{Pre-mortem} & \textbf{Freq} & \textbf{Post-mortem} & \textbf{Freq} & \textbf{Pre-mortem} & \textbf{Freq} & \textbf{Post-mortem} & \textbf{Freq} \\
    \cmidrule(lr{.75em}){1-4} \cmidrule(lr{.75em}){5-8}
    love & 126,197 & love  & 231,921 & love love   & 19,665 & love miss   & 30,605 \\
	hey  & 95,358  & miss  & 227,834 & love ya     & 13,220 & miss love   & 18,864 \\
	lol  & 83,993  & know  & 111,612 & let know    & 7,130  & love love   & 12,696 \\
	just & 73,453  & just  & 108,783 & heart heart & 7,106  & love ya     & 12,348 \\
	im   & 64,683  & like  & 71,449  & just wanted & 6,676  & just wanted & 12,095 \\

  \bottomrule
\end{tabular}
\end{center}
\end{table}

\begin{table}%
\begin{center}
\caption{Mann-Whitney $U$-tests results and descriptive statistics of pre- and post-mortem comments. Only the differences with at least small effect sizes ($|d|>0.2$) are reported. *$p < 0.0001$}
\label{tab:3}
\begin{tabular}{cSScS}
    
	\textbf{LIWC Metric} & \textbf{Pre-mortem} & \textbf{Post-mortem} & \textbf{$p$} & \multicolumn{1}{c}{\textbf{$d$}} \\
    \midrule
    Sadness               & 0.007      & 0.04        & * & -0.682 \\
    Second person pronoun & 0.051      & 0.093       & * & -0.549 \\
    Word count            & 0.232      & 0.488       & * & -0.523 \\
    Social processes      & 0.14       & 0.191       & * & -0.389 \\
    First person pronoun  & 0.067      & 0.092       & * & -0.361 \\
    Negative emotion      & 0.026      & 0.05        & * & -0.352 \\
    Present tense         & 0.104      & 0.124       & * & -0.215 \\
    \bottomrule
\end{tabular}
\end{center}
\end{table}

\subsection{Analysis of Post-Mortem Language}
Our prior analysis identified significant differences between pre- and post-mortem content. For example, the average word count of post-mortem comments ($\mu = 48.79$) is significantly greater than pre-mortem comments ($\mu = 23.22$;  Mann-Whitney $U = 5.9 \times 10^{10}$,  $p < 0.001$), in line with what others have found on smaller datasets \cite{getty_i_2011}. We also identified language differences between pre- and post-mortem content. In order to provide a rich sense of the dataset, we have provided the top 5 most frequently occurring unigrams and bigrams in Table \ref{tab:2}.




In terms of linguistic characteristics, significant differences between the pre- and post-mortem content also exist. We analyzed linguistic characteristics using ``Linguistic Inquiry and Word Count,'' (LIWC)\footnote{http://liwc.wpengine.com/how-it-works/} , a common language analysis package that provides dictionaries for parts of speech and punctuation, as well as psychosocial and social processes. For each body of text, LIWC provides a score between 0-100 (normalized to 0-1 in our analysis), indicating the proportion of words in the text contained in the dictionary for each given category. Descriptive statistics, $p$-values (corrected using Holm-Bonferroni method), and effect sizes (Cohen's $d$) are shown in Table \ref{tab:3}.
\subsection{Classification}
\label{sec:prior}
In our previous work \cite{ICWSM1817819}, we reported on some initial success in classifying the mortality of profiles. Our unit of analysis was the concatenated comments of each profile. We used an $n$-gram-based logistic regression classifier and achieved an F1 score of $0.837$ with 10-fold cross validation. 

While we achieved a promising result from profile-level classification, our approach was  a limited first step. We did not look into the informative features that signal pre- or post-mortem status, which would provide greater insight into the characteristics of post-mortem language in our training data. Other natural next steps of this previous work were to try other classifiers and to classify mortality on a more granular level, i.e. on the level of individual comments. In this work, we address all of these issues, but also show we can quickly classify mortality with a small amount of post-mortem content, and both profile- and comment-level classifiers work well across different platforms.

\section{Classifying Post-mortem Content}
\label{sec:classify}

In this section, building on our initial investigation into post-mortem language, we present two machine learning classification tasks: to classify (1) if a profile is pre-mortem or post-mortem (i.e., whether the profile owner is alive or dead), and (2) whether a single comment was authored before or after its recipient's death. We discuss the classification methods used in each task and their associated classification results and informative features. Finally, we discuss how early we can determine the post-mortem status of a profile.

\subsection{Profile Level Classification}
\label{sec:profile_classify}
We extended our previous work on classifying post-mortem profiles by experimenting with more classifiers and features. We first developed a simple baseline rule-based classifier: Classify a comment as post-mortem if the comment contains the word ``rip,'' and pre-mortem otherwise. We then implemented four commonly used text classifiers to compare with our baseline: Multinomial Naive Bayes (NB), Logistic Regression (LR), Linear SVM (SVM), and Boosted Trees. NB, LR, and SVM were implemented with the Python machine learning library scikit-learn; Boosted Trees was implemented with the XGBoost system described in \cite{chen_xgboost:_2016}. 

For these four classifiers, we compared three different feature sets: 
\begin{enumerate}
  \item $n$-gram features ($n = 1,2,3$) with TF-IDF weights; 
  \item features derived from computational linguistic tools (CLT): style, topic, and sentiment measures from LIWC \cite{tausczik_psychological_2010}, Empath's default categories  \cite{fast_empath:_2016} and VADER \cite{hutto_vader:_2014} ; and 
  \item the combination of (1) and (2). 
\end{enumerate}
    
For feature sets (1) and (3), we additionally conducted $\chi^2$ feature selection due to the large number of features derived from our $n$-gram model. The number of features we tried ranged from 1,000 to 10,000 in 1,000 intervals. We removed VADER's \emph{compound} metric from our feature sets because its value can be negative, which is incompatible with our $\chi^2$ feature selection, and the compound metric's collinearity with VADER's \emph{pos}, \emph{neg}, and \emph{neu} makes it a redundant feature. We removed all stopwords and used GridSearch to tune the classifier parameters (e.g., regularization, tolerance, max tree depth). We evaluated our classifiers using F1 scores and 10-fold cross validation.

\subsubsection{Findings}
Table \ref{tab:profile} shows the classification result of each classifier with each feature set. We only report the best-performing feature selection results for the sake of brevity. The baseline classifier achieved an F1 score of 0.74 and a recall of 0.78. In other words, there was at least one ``rip'' posted to 78\%  of the post-mortem profiles, which is consistent with the norm of how people speak to the deceased offline. Overall, XGBoost with 4,000 features from feature set (3)---$n$-grams and linguistic tools features combined---had the best performance, with an F1 score of 0.882. We conducted paired $t$-tests to compare the mean cross-validation scores of the best-performing classifier and the other classifiers, and found with $p < 0.05$, XGBoost with 4,000 features from feature set (3) performed better than the other classifiers except for one -- XGBoost with $n$-grams only. In other words, the inclusion of CLT features did \emph{not} significantly improve the performance of our best-performing classifier.
%
%
%
%

\begin{table}%
\begin{center}
\caption{Classification metrics of profile level classifiers. \\ Classifier with the best cross-validation F1 score is shown in bold.}
\label{tab:profile}
\begin{tabular}{cccccc}
    
	 &  & Accuracy & F1 & Precision & Recall \\
    \midrule
    & Baseline & 0.696 & 0.736 & 0.697 & 0.778 \\
    \midrule
    \multirow{4}{*}{$n$-gram} & NB & 0.835 & 0.837 & 0.901 & 0.782 \\
	& LR       & 0.856 & 0.858 & 0.946 & 0.785 \\
	& SVM      & 0.862 & 0.866 & 0.952 & 0.794 \\
	& XGBoost  & 0.876 & 0.881 & 0.942 & 0.827 \\
    \midrule
    \multirow{4}{*}{CLT} & NB       & 0.593 & 0.72  & 0.578 & 0.953 \\
	& LR       & 0.75  & 0.769 & 0.775 & 0.764 \\
	& SVM      & 0.789 & 0.793 & 0.846 & 0.747 \\
	& XGBoost  & 0.821 & 0.828 & 0.884 & 0.779 \\
    \midrule
    \multirow{4}{*}{$n$-gram + CLT}& NB       & 0.846 & 0.84  & 0.961 & 0.746 \\
	& LR       & 0.856 & 0.858 & 0.939 & 0.79  \\
	& SVM      & 0.865 & 0.865 & 0.952 & 0.793 \\
	& \textbf{XGBoost}  & \textbf{0.874} & \textbf{0.882} & \textbf{0.942}  & \textbf{0.829} \\
    
    \bottomrule
\end{tabular}
\end{center}
\end{table}

\begin{table}%
\small
\begin{center}
\caption{Ten most informative features of profile level classifiers. \\ *Metric from LIWC. $^\dagger$Metric from VADER.}
\label{tab:profile_feature}
\def\tabularxcolumn#1{m{#1}}
\rowcolors{2}{gray!25}{white}
\begin{tabularx}{\linewidth}{YYYYYY}
	\rowcolor{white}
    \multicolumn{2}{c}{$n$-gram} & \multicolumn{2}{c}{CLT} & \multicolumn{2}{c}{$n$-gram + CLT}\\
    \cmidrule(lr{.75em}){1-2} \cmidrule(lr{.75em}){3-4} \cmidrule(lr{.75em}){5-6}
    \rowcolor{white}
	\textbf{Pre-mortem} & \textbf{Post-mortem} & \textbf{Pre-mortem} & \textbf{Post-mortem} & \textbf{Pre-mortem} & \textbf{Post-mortem} \\
    \midrule
    
hey     & miss      & Question marks*      & Sadness words*       & hey             & rip       \\
wait    & rip       & \# of unique words*  & Personal pronouns*   & Question marks* & miss      \\
need    & wish      & Assent words*        & Negative emotion*    & wait            & wish      \\
hang    & missed    & Swear words*         & Negative$^\dagger$            & need            & gone      \\
yo      & gone      & Anger words*         & Cognitive processes* & hang            & missed    \\
soon    & know      & Auxiliary verbs*     & Common verbs*        & ya              & know      \\
seen    & thinking  & Impersonal pronouns* & Function words*      & soon            & thinking  \\
come    & heaven    & Tentativeness words* & Total pronouns*      & seen            & heaven    \\
ooooooo & miss love & Exclamation marks*   & Quantifiers*         & come            & miss love \\
home    & forget    & Positive emotion*    & Affective processes* & yo              & forget   \\
	\bottomrule
	
\end{tabularx}
\end{center}
\end{table}

Extending the profile-level classification results from our previous work, we took a closer look at the classification task and analyze the informative features. Table \ref{tab:profile_feature} shows the ten most informative features for pre-mortem and post-mortem classes from each feature set of the profile-level classifiers. Informative features from feature set (2), which only includes linguistic measures, showed that higher use of sadness words, personal pronouns, negative emotion, cognitive processes, common verbs, function words, total pronouns, quantifiers, and affective processes, is associated with higher likelihood of being post-mortem; question marks, proportion of unique words, assent, swear words, angry words, auxiliary verbs, impersonal pronouns, tentativeness words, exclamation marks, and positive emotion indicate higher likelihood of being pre-mortem.
    
Informative features from feature set (1) showed that words like ``miss,'' ``rip,'' and ``gone'' were associated with higher likelihood of being post-mortem, while words like ``hey,'' ``wait,'' ``yo'' were associated with higher likelihood of being pre-mortem. Informative features from feature set (3) contained mostly $n$-gram features, indicating the presence of particular words or phrases overwhelms the metrics from computational linguistic tools, which confirmed the result of our statistical significance test.

\subsection{Comment Level Classification}
\label{sec:comment_classify}
Building on our success with profile-level classification, we next attempted more granular classification. In many cases, post-mortem content is encountered independent of the deceased's profile (e.g., through a news feed \cite{brubaker_beyond_2013}, news article \cite{gach_control_2017}, or group \citep{getty_i_2011,marwick_there_2012}) but is still upsetting to those who see it. In order to account for this scenario, we tried to classify post-mortem content at the level of a single comment. In the comment level model, we sought to classify whether a single comment was posted pre-mortem or post-mortem based on the language of the comment. We used the same classification methods used in the profile-level task described in Section \ref{sec:profile_classify}. 

\begin{table}%
\begin{center}
\caption{Classification metrics of comment level classifiers. \\ Classifier with the cross-validation best F1 score is shown in bold.}
\label{tab:comment}
\begin{tabular}{cccccc}
    
	 &  & Accuracy & F1 & Precision & Recall \\
    \midrule
    & Baseline & 0.662 & 0.204 & 0.892 & 0.115 \\
    \midrule
    \multirow{4}{*}{$n$-gram} & NB       & 0.881 & 0.836 & 0.87  & 0.804 \\
    & LR       & 0.898 & 0.858 & 0.898 & 0.822 \\
    & SVM      & 0.897 & 0.857 & 0.898 & 0.819 \\
    & XGBoost  & 0.867 & 0.812 & 0.869 & 0.761 \\
    \midrule
    \multirow{4}{*}{CLT} & NB       & 0.78  & 0.633 & 0.848 & 0.505 \\
    & LR       & 0.833 & 0.77  & 0.81  & 0.734 \\
    & SVM      & 0.847 & 0.79  & 0.818 & 0.763 \\
    & XGBoost  & 0.871 & 0.82  & 0.862 & 0.781 \\
    \midrule
    \multirow{4}{*}{$n$-gram + CLT}& NB       & 0.888 & 0.848 & 0.87  & 0.827 \\
    & LR       & 0.886 & 0.843 & 0.881 & 0.808 \\
    & \textbf{SVM}      & \textbf{0.903} & \textbf{0.865} & \textbf{0.901} & \textbf{0.833} \\
    & XGBoost  & 0.884 & 0.838 & 0.885 & 0.795 \\
    
    \bottomrule
\end{tabular}
\end{center}
\end{table}

\subsubsection{Findings}
The classification results can be seen in Table \ref{tab:comment}. The baseline model did not perform well, with an F1 score of only $0.2$. The baseline classifier had a precision of $0.89$, but only a recall of 0.11, indicating that while ``rip'' is a strong indicator of post-mortem comments, it only showed up in 11\% of the post-mortem comments in our dataset. Overall, SVM with all features from feature set (3) had the best performance, with an F1 score of $0.865$. We further conducted paired $t$-tests to compare the mean cross-validation scores of the best-performing classifier and the other classifiers, and with $p < 0.05$, we found SVM with all features from feature set (3) indeed performed better than all the other classifiers. This result, compared with that of profile-level classifiers, indicated CLT features helped comment-level classifiers much more than profile-level classifiers.

We present the 10 most informative features based on feature coefficients for pre-mortem and post-mortem classes from each feature set in Table \ref{tab:comment_feature}. Informative features from feature set (2), which only includes linguistic measures, showed that higher use of sadness words, quantifiers, certainty words, time, negative emotion, common verbs, religion words, total second-person pronouns, insights, and cognitive processes, are associated with higher likelihood of being post-mortem; assent, proportion of unique words, question marks, auxiliary verbs, tentativeness, anger, swear words, positive emotion, impersonal pronouns, and exclusion indicate higher likelihood of being pre-mortem. 

\begin{table}%
\small
\begin{center}
\caption{Ten most informative features of comment level classifiers. \\ *Metric from LIWC.}
\label{tab:comment_feature}
\def\tabularxcolumn#1{m{#1}}
\rowcolors{2}{gray!25}{white}
\begin{tabularx}{\linewidth}{YYYYYY}
	\rowcolor{white}
    \multicolumn{2}{c}{$n$-gram} & \multicolumn{2}{c}{CLT} & \multicolumn{2}{c}{$n$-gram + CLT}\\
    \cmidrule(lr{.75em}){1-2} \cmidrule(lr{.75em}){3-4} \cmidrule(lr{.75em}){5-6}
    \rowcolor{white}
	\textbf{Pre-mortem} & \textbf{Post-mortem} & \textbf{Pre-mortem} & \textbf{Post-mortem} & \textbf{Pre-mortem} & \textbf{Post-mortem} \\
    \midrule
    
need hang      & rip          & Assent words*        & Sadness words*       & Assent words*        & rip            \\
yeah           & heaven       & \# of unique words*  & Quantifiers*         & lol                  & miss           \\
lol            & missin       & Question marks*      & Certainty*           & \# of unique words*  & Sadness words* \\
stay safe      & better place & Auxiliary verbs*     & Time words*          & Question marks*      & wish           \\
aww            & grave        & Tentativeness words* & Negative emotion*    & haha                 & thinking       \\
feeling better & forgotten    & Anger words*         & Common verbs*        & Tentativeness words* & Quantifiers*   \\
haha           & wish         & Swear words*         & Religion*            & yeah                 & day            \\
havent talked  & memorial     & Positive emotion*    & 2nd person pronouns* & hang                 & missed         \\
hang soon      & missing      & Impersonal pronouns* & Insight words*       & Auxiliary verbs*     & gone           \\
love song      & thinking     & Exclusion words*     & Cognitive processes* & Anger words*         & love miss \\
	\bottomrule
	
\end{tabularx}
\end{center}
\end{table}

Informative features from feature set (1) show that the words like ``rip'' and ``heaven'' were associated with post-mortem content, while words like ``hang,'' ``stay safe,'' and ``lol'' were associated with pre-mortem content. Informative features for post-mortem from feature set (3) contained mostly n-grams, indicating the presence of particular words or phrases were more important than linguistic features. For pre-mortem comments, linguistic features from LIWC were overall more important than particular words.

\subsection{How Early Can We Predict Post-Mortem Profiles?}
Given our success at identifying mortality at the level of individual comments, we next sought to determine how much data our classifier needed to observe in order to make an accurate classification. There are many scenarios in which early classification is important. System designers, for example, may need to sensitize their interactions or provide death-specific functionality within days or even hours of the death. Using the profile level model, we looked at how fast we could classify post-mortem profiles. We randomly selected 20\% of the profiles as test profiles, and the rest 80\% as training profiles. We then removed all comments that were made post-mortem in the test profiles, and then used the best-performing profile-level model classifier to classify test profiles as we incrementally added post-mortem messages to each profile one by one in chronological order. We measured the speed at which we can determine post-mortem profiles in terms of the number of post-mortem comments. Using the comment timestamps, we also measured the speed in terms of the actual time after the first comment was made.
%
%

\begin{figure}
%
%
  \includegraphics[scale=0.7]{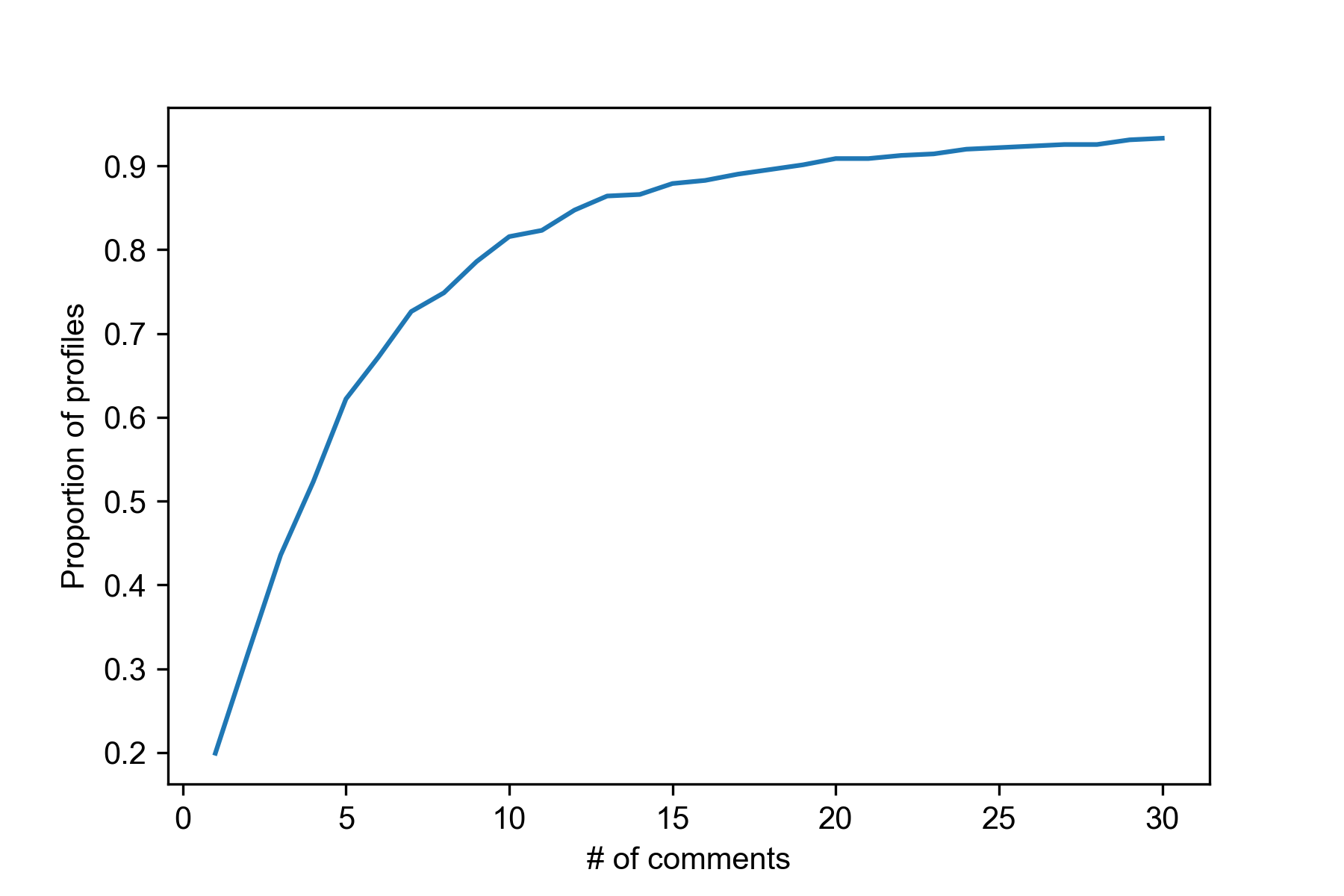}
  \caption{Percentage of profiles classified vs. number of post-mortem comments provided.}
  \label{fig:1}
\end{figure}

We were able to determine post-mortem profiles quickly (see Fig. \ref{fig:1}). We were able to determine 19.9\% of the post-mortem profiles after the first post-mortem comment was made, 52.3\% after the first four post-mortem comment were made, and 90.1\% after the first nineteen comments were made. In our dataset, this result means that we were able to classify 31.8\% of the post-mortem profiles on the day of death, 52.3\% within the first day after death, and 90.1\% within ten days after death.

\subsection{Error Analysis}
Despite our classifiers' good performance, the sensitivity of death encouraged us to qualitatively investigate misclassifications. To demonstrate challenges faced in classification we provide and discuss example pieces of data. In selecting these examples, we only share messages that cannot be  traced back to the original content through web search, and we have omitted names to protect the survivors' privacy.

In most cases, false positives (F/P; pre-mortem content classified as post-mortem) occurred when comment content included emotional language often seen post-mortem, such as ``love'', ``miss'', and ``sorry.'' Short messages (e.g., ``I love you too'') can be ambiguous as they lack context, but longer messages can present challenges when their length results in language that is often associated with post-mortem content. For example, the following comment included words like ``heaven'', LIWC's social process words (e.g., ``sister''), as well extensive use of pronouns:

\begin{quote}
...I just wanted to let you know that I'm your big sister period. The one and only and I'll always be your big sister. I never ever want you to feel like you can't come to me and talk to me... I promise to the lord in heaven that as long as I'm alive, rain or shine... I'll always and forevermore be here for you...
\end{quote}

False negatives (F/N), while sharing many of the same challenges as F/Ps, have some unique qualities as well. We briefly mention three.

Some post-mortem comments resemble pre-mortem content. Previous work has noted the role of memorial profiles in ``continuing bonds'' \citep{klass_continuing_1996} with deceased loved ones, maintaining connections---and conversations---in line with those that existed pre-mortem \citep{getty_i_2011,degroot_maintaining_2012,brubaker_we_2011}. Consider the following comment:
%
%

\begin{quote}
Hey bro. Spring seems to be here. Now time for new life to begin. I [hope] you are doing ok. I['m] just eating a bagel. I keep telling myself you guys are in a better place.
\end{quote}

Casual, and conversational language is not uncommon, but our classifiers can get confused when users author comments that perform an ongoing relationship. 

Second, it is not uncommon for people to include quotes, scriptures, poems, or musical lyrics in their comments. The text of comments, such as the lyrics to a Green Day song one user posted, present a significant challenge for classification. One possible approach would be to computationally identify and then remove lyrics, similar to Kivran-Swaine et al. \cite{kivran-swaine_understanding_2014} .

Finally, in some cases, even though comments were added post-mortem, they were clearly written by a user who assumed the account holder was alive. For example:

\begin{quote}
Hey, how's it all goin? Wats new? I'm sorry to hear bout your sister. Much love! 
\end{quote}

Understandably, our comment-level classifier cannot account for these comments. However, their presence highlights the need for both comment-level and profile-level classification, as well as the role that classification can play in more clearly communicated the mortal status of social media users. 

\section{Testing Classifiers}
As a final step, to see how well our classifiers can be generalized to data from other platforms and time frames, we tested our classifiers on two additional datasets: Facebook Memorial Groups, and Newspaper Obituaries. These datasets contain post-mortem content that comes from a much more recent timeframe and represents both casual and formal writing about death. Examining the performance of classifiers trained on a nearly decade-old dataset from MySpace on these datasets will shed light on whether and how memorial language has changed over time, and what it means for today's social media. We describe each dataset and their classification results below, followed by a discussion of the results from both datasets.

\subsection{Facebook Memorial Groups}
\label{sec:fb}
Our Facebook Memorial Groups dataset contained messages posted in groups created by friends and family to memorialize the deceased. Given that these groups are created post-mortem, this dataset does not include any pre-mortem content posted by either the deceased or the survivors. We chose to validate our classifier on this dataset as it is similar to the post-mortem MySpace data on which we trained our classifiers---they are both data from social media in which friends are memorializing a loved one. However, testing on this dataset allows us to validate our classifiers on a different social media platform, and with more recent content that spans a larger period of time. 
%
%
%
%

\subsubsection{Data Collection}
We started data collection by manually identifying Facebook Memorial Groups. Following Marwick \& Ellison \cite{marwick_there_2012}, we identified  116 groups by searching for public Facebook Groups with the keywords ``RIP'' (and its variants) and ``in memory of'' using the Facebook search bar. We excluded groups dedicated to celebrities or groups that are not created to memorialize deceased people (e.g., ``RIP Harambe''). After manually verifying each group, we used the Facebook Graph API to collect all messages from these groups, including posts and any replies. Our final dataset contained 90,100 individual messages posted to Facebook between November 2006 to March 2018. 32,196 of these messages were posts and 57,904 were replies.

\subsubsection{Results}
Using the best-performing comment-level and profile-level classifiers trained with MySpace data, we classified mortality at both the level of individual messages and the memorial group as a whole. 
Since this dataset does not contain negatives (i.e., pre-mortem) and precision and F1 score are not defined in this case, we only report the recall score---how much post-mortem content our classifiers could identify. 

Our best-performing comment-level classifier (SVM, $n$-gram + CLT, all features) achieved a recall of 0.457 on classifying all individual messages (posts \& replies), 0.688 on a posts-only subset, and 0.329 on a reply-only subset. Our best-performing profile level classifier (XGBoost, $n$-gram + CLT, 4,000 features) achieved a recall of 0.871, on par with our validation score on the MySpace dataset.


\subsection{Newspaper Obituaries}
Our dataset of newspaper obituaries consisted obituaries authored from across the US and published by local newspapers and hosted by Legacy.com. These obituaries are conventional, but Legacy.com does include a guestbook feature which contains messages posted in response to the obituaries. Compared to our training data from MySpace, this dataset was not collected from social media and consists of the more formal language  seen in public obituaries.

\subsubsection{Data Collection}
For this dataset, we scraped obituaries posted on U.S. newspapers and their guestbooks that contains messages posted to these obituaries. The obituaries and messages were scraped from Legacy.com using the web tool import.io. This dataset consisted of 4,659 obituaries published during April 2017, as well as all 4,537 guestbook messages posted to these obituaries from April to September 2017.

\subsubsection{Results}
We used the same set of classifiers from Section \ref{sec:fb} to classify mortality of the obituaries, individual guestbook messages, and guestbooks as a whole. 
Our best-performing comment-level classifier (SVM, $n$-gram + CLT, all features) achieved a recall of 0.797 on obituaries, and 0.864 on individual guestbook comments. Our best-performing profile level classifier (XGBoost, $n$-gram + CLT, 4,000 features) achieved a recall of 0.46 on guestbooks.

\subsection{Error Analysis}
We present a brief error analysis on our test sets. As our datasets do not contain negatives, all misclassifications here are false negatives (F/N). Here, again, we only provide quotes that cannot be easily traced back to the original content through web search, and we have omitted names to protect the survivors' privacy.

First, similar to what we presented in the previous error analysis, some post-mortem messages resemble pre-mortem ones, which do not use words that clearly indicate death. Take, for example, the following message posted to an obituary guestbook:
%
%
%
%
\begin{quote}
%
%
%
[Name] is such a pleasant person. We didn't see her often but when we did she always welcomed you with a caring and warm heart.
\end{quote}


Second, by looking at the features' trained weights, we found some condolence words commonly addressed to the survivors were not associated with death in our training dataset. Instead, words like ``sorry'' and ``prayer'' had negative weights in our trained classifiers and were therefore associated with pre-mortem content, likely due to their occurrence in ongoing pre-mortem conversations. This association resulted in misclassification. For example, ``sorry'' and ``prayer'' are the only indicators of death in the following message, which confused our classifier:
%
%
%
%
%
%
%
%
\begin{quote}
%
%
%
So sorry to hear this news. My prayers go out to his wife and son [name]. [Name] was a great neighbor and friend of my father's for a number of years.
\end{quote}

Finally, it is surprising that the profile-level classifier in the Newspaper Obituaries dataset at the guestbook-level achieved a worse result than the message-level classifier, given that there is more context on the level of guestbooks as a whole. These results are also the inverse of what we observed with the MySpace data where the profile-level classifier performed better than the comment-level classifier. However, we did find that the Neutral metric from VADER had a negative weight, meaning it was informative of pre-mortem content. For this reason, we speculate the low recall of guestbook-level classification is a result of the high Neutral score from VADER caused by message concatenation.
%
%
%
%
%
%
%
%

Overall, the classification results from the test set are comparable with the validation scores from the MySpace dataset, which suggests that our classifiers are generalizable to other platforms as well as more recent data. More importantly, this result also suggests language of death and bereavement may be consistent across time, contexts, and platforms, which we explore in detail in our discussion.
%
%
%
%
%
%

\section{Discussion}
The results of the classification tasks show that post-mortem profiles and memorial comments can be identified with a high degree of accuracy. In both classification tasks, the feature set that combined $n$-grams and linguistic measures had the best performance. This result indicates that both specific words and linguistic style are important for identifying post-mortem status. 

In the best-performing feature set for both tasks, informative features include typical memorial language, such as ``rip,'' ``miss,'' and ``gone.'' These words suggest that offline mourning norms are also present in an online setting (c.f., \citep{getty_i_2011}); they also suggest that our classifiers might generalize to other platforms, which we were able to confirm when we successfully tested our classifiers on other datasets. While informative features at the profile level almost exclusively include $n$-grams, the most informative features at the comment-level also include sentiment and linguistic style metrics like LIWC's sadness and quantifier metrics. We also observed that these metrics significantly improved the performance of our comment-level classifiers but not profile-level classifiers. The importance of these metrics indicates that, in the absence of the context that other comments provide, linguistic style and sentiment become good predictors of memorial content. 

Profile-level informative features for pre-mortem status include ``hey,'' ``yo,'' ``need,'' ``hang'' and question marks. These features suggest that greeting words, and words that indicate future events are strong indicators of pre-mortem status. For comment-level, informative $n$-grams include ``lol,'' ``haha,'' and ``yeah,'' which indicate ongoing conversations. Linguistic features, including assent words, question marks, and tentativeness, also indicate ongoing conversations and question-answering. Once again, comment-level informative features for pre-mortem status contain a large proportion of linguistic style and sentiment features (6 out of the top 10), which means they are strong predictors in the absence of other comments to provide context.

The good results we achieved when testing our classifiers on data from other platforms have important implications for understanding post-mortem content. One might reasonably question the generalizability of our classifiers to other contexts or types of data---we did as well. We hypothesized that the classifiers would have decreased performance on the Facebook dataset, and even more so on the newspaper obituary dataset. The formal language used in obituaries and eulogies, for example, is distinct from the casual and conversational language in our training dataset. Likewise, our training data are drawn from MySpace in 2010, and changes in linguistic practices since then could have presented challenges. However, our classifiers still performed well on data from other platforms and with more recent data. While Gibbs et al. have argued that social media design and use can result in distinctive ``platform vernacular'' \cite{gibbs_funeral_2015}, our classifiers' good performance suggests that the vernacular of post-mortem messages has some consistency across platforms, time, and genre, and the language around death and bereavement might be more universal and enduring than we thought.

\section{Sensitive Classification in Context}
Classifying post-mortem content, and mortality more broadly, is important for both platform designers and data scientists, but requires additional considerations depending on the context. 

In the context of social media platforms, classification can enable designers to shape when and how people encounter post-mortem content, as well as how that content is presented. For example, prior research has shown that unexpected encounters with post-mortem content can be distressing, especially when it is encountered in a stream of typical (and often casual) social media content \cite{brubaker_beyond_2013}. Likewise, the broadly public nature of social media can also result in conflicts when strangers interact around public deaths \cite{gach_control_2017}. As social media extends the participation and witnessing of memorialization and grief \cite{walter_does_2012}, identifying post-mortem content can help platform designers shape where and how people encounter content, as well as where memorializing practices occur. For example, while many people want deceased loved ones to remain part of their social networks \cite{pennington_you_2013}, it may be inappropriate to recommend they be invited to an event or included in a group chat. Profile-level classification can provide ways to distinguish between living users and those who have died.
%
%
%
%

Sensitivity in the context of social media platforms also extends to time. There are scenarios in which it may be desirable to control access to a profile or shape how sensitive information is shared immediately following a death. For example, designers may want to limit access to a publicly visible profile representing the deceased to prevent trolling or unwelcome strangers \cite{marwick_there_2012,degroot_for_2014}. In other cases, it might be appropriate to restrict information about the death, giving people the opportunity to inform friends and loved ones in more face to face or synchronous settings. In these cases, designers may find classifiers that are able to quickly determine mortality useful. 

Beyond platform design, data scientists should account for mortality in their work, especially when making use of user-generated content and social data. Here we echo an argument we first encountered when Geiger and Halfaker \cite{geiger_using_2013} developed techniques to more accurately understand the meaning of our social data (in their case, examining user sessions on Wikipedia instead of just activity logs). We imagine that in most cases scientists will want to avoid unintentionally including the dead. Ma et al. \cite{ma_write_2017} provide us with one example of how accounting for mortality can improve and clarify results, and implicitly, how ignoring mortality can bias data science models. Moreover, developing techniques to account for mortality will benefit scholarship with an explicit post-mortem focus, as well as scholarship that seeks to compare pre- vs. post-mortem content.
%
%

Finally, when classifying content content related to death in both of these contexts, it will be important to look beyond classification performance and instead focus on specific types of precision---a clear next step for future research. While our classifiers performed well, the sensitivity death can require suggests that in certain circumstances, precision should be prioritized, even at the expense of overall performance.  In practice, we recommend researchers and practitioners to consider the stakeholders and consequences of the specific task in order to determine the appropriate precision-recall balance. For example, the designer of a social media platform would likely want to avoid false-positives, rather than incorrectly classify some an account as post-mortem. Conversely, a computational social scientist may want to avoid false-negatives, electing to remove users who may be alive from their dataset to ensure that post-mortem accounts are not included. 

We would caution against simplistic or out-of-the-box use of a mortality and memorial content classifiers. This is especially true in the case of platform design where the consequences of misclassification can be extreme, and it is important for assurances to be made around the accuracy of the classifier. Designers cannot take a classifier such as ours and apply it in black-and-white ways. Instead, they must think carefully about the context in which it is being used and the ways it will shape the interactions they design. Given the sensitive nature of this problem space, we suggest that, even though a classifier might label a profile as post-mortem, system designers are under no obligation to treat it literally as such. Conservative interpretations, such as ``possibly post-mortem'', can still provide designers a signal with which to sensitize their designs, without, for example, making publicly visible changes to an account or profile (e.g., deleting or memorializing an account).
%
%


We would also encourage designers to not limit themselves to considering mortality at the level of users and profiles. While our focus on post-mortem profiles and their comments based was initially motivated by the extensive literature documenting memorializing practices and linguistic differences pre- and post-mortem, one potential benefit of our approach is the ability to detect post-mortem comments independent of the profile. Prior research has found that unexpected encounters with post-mortem content are the most distressing \citep{brubaker_beyond_2013}. These encounters are often the result of messages that propagate through ``social awareness streams'' \cite{Naaman:2010:RMM:1718918.1718953} like Facebook's News Feed, but lose the memorial context of the post-mortem profile in the process. One can imagine how memorial content detection might have also prevented Meyer's distressing experience shared at the top of the paper. The high-accuracy of our comment-level approach may allow platform designers detect potentially distressing messages and design sensitive interactions with mortality in mind.

In all of the aforementioned cases, it is worth acknowledging that even as classification provides designers with the ability to detect mortality and memorial content, the ways these tools are used can have profound impacts on death and memorialization---not only in our platforms, but for our culture as well. One could imagine that if Facebook had the ability to easily detect post-mortem profiles years ago, they might have simply deleted them out of caution. Such a choice, however, would have precluded the development of the common memorial spaces that many people cherish today. With this in mind, we would recommend that designers solicit the perspectives of domain experts when deciding how to utilize classifiers in sensitive contexts. Social scientists, mental health providers, and chaplains with expertise in death and bereavement are well positioned to help designers understand the risks and trade-offs. Likewise, designers must also consider how to solicit feedback from the people who use our platforms. How to conduct human-centered evaluation of classifiers (without the context of a specific interaction) is a new and open problem for CSCW, but there is some emerging work that is trying to approach the use of classifiers in value-sensitive and human-centered ways \cite{zhu_value_2018}. 
%
%



Finally, we recommend researchers and practitioners in other areas of CSCW to consider mortality in their work, especially in cases where longitudinal effects are involved. For example, mortality may be an important factor in using social media data to study longitudinal effects of infectious diseases or mental illnesses, as the deceased would not be ``active'' throughout the entire time period and thus bias the results. Practitioners who use machine learning models to predict future events, such as outcomes of political elections, may also benefit from considering mortality, as a deceased person cannot impact the future outcome of an event.

\section{Conclusion}
In this paper, we reported on accurate machine learning classifiers for both post-mortem profiles and post-mortem comments and discussed informative linguistic features for classifying mortality. We further showed that we could identify deceased users soon after their death. Finally, we tested our trained classifiers on data from two other platforms and also had good result.
%
%

The major contributions of this work lie in introducing an automated and accurate way to identify post-mortem content that can be incorporated into design processes and data science models, while current platform implementations require significant human involvement. This work also challenges the assumption often made by design frameworks and data science models that people behind the data are alive, which may lead to distressing design choices and inaccurate data science models.
\begin{acks}
We would like to thank the reviewers for their significant time and care in improving this paper. We would also like to thank Amjad Alharbi, Casey Fiesler, Kyle Frye, Brian Keegan, Michael Paul, and Danielle Szafir for their invaluable help and feedback on this paper.
\end{acks}

\bibliographystyle{ACM-Reference-Format}
\bibliography{ripgenre}

\end{document}